# Filamentary Switching: Synaptic Plasticity through Device Volatility

Selina La Barbera, Dominique Vuillaume, and Fabien Alibart[*]

*Institut d'Electronique, Microelectronique et Nanotechnologies, CNRS,*
*Avenue Poincaré, 59652 Villeneuve d'Ascq, France*

**ABSTRACT**. Replicating the computational functionalities and performances of the brain remains one of the biggest challenges for the future of information and communication technologies. Such an ambitious goal requires research efforts from the architecture level to the basic device level (i.e., investigating the opportunities offered by emerging nanotechnologies to build such systems). Nanodevices, or, more precisely, memory or memristive devices, have been proposed for the implementation of synaptic functions, offering the required features and integration in a single component. In this paper, we demonstrate that the basic physics involved in the filamentary switching of electrochemical metallization cells can reproduce important biological synaptic functions that are key mechanisms for information processing and storage. The transition from short- to long-term plasticity has been reported as a direct consequence of filament growth (i.e., increased conductance) in filamentary memory devices. In this paper, we show that a more complex filament shape, such as dendritic paths of variable density and width, can permit the short- and long-term processes to be controlled independently. Our solid-state device is strongly analogous to biological synapses, as indicated by the interpretation of the results from the framework of a phenomenological model developed for biological synapses. We describe a single memristive element containing a rich panel of features, which will be of benefit to future neuromorphic hardware systems.

Massive amounts of heterogeneous data are generated each day in our society. In this context, computing systems face important challenges in providing suitable solutions for information processing. Saturation of conventional computer performances due to material issues (*i.e.*, clock frequency and energy limitations) and more fundamental constraints inherent in the Von Neumann bottleneck have forced researchers to investigate new computing paradigms that will allow for more powerful systems. The bio-inspired approach (or, more precisely, neuromorphic engineering) is a promising direction for such an objective. Recent breakthroughs at the system,[1] circuit,[2] and device levels[3] are very encouraging indicators for the development of computing systems that can replicate the brain's performances in tasks such as recognition, mining, and synthesis.[4] To achieve such an ambitious goal, research efforts are needed for understanding the computing principles of biological systems, elucidating how spike-coding information is computed and stored in neuron and synapse assemblies, and exploring neuromorphic approaches that define hardware functionalities, performances, and integration requirements. Emerging nanotechnologies could play a major role in this context, by offering devices with attractive bio-inspired functionalities and associated performances that would ensure the future development of neuromorphic hardware. Some studies have investigated the possibility of implementing neurons in nano- scale devices.[5,6] Most of these efforts have been devoted to the realization of synaptic elements with emerging memory devices, such as RRAM technologies, with the goals of matching the critical integration density of the synaptic connections[7] and replicating the synaptic plasticity mechanisms that correspond to the modification of synaptic conductance during learning and computing. Indeed, modification of the synaptic weight as a function of neuronal activity (*i.e.*, spiking activity) is widely recognized as a key mechanism for the processing and storage of information in neural networks. Plasticity mechanisms are commonly categorized as short- and long-term plasticity (STP and LTP, respectively).

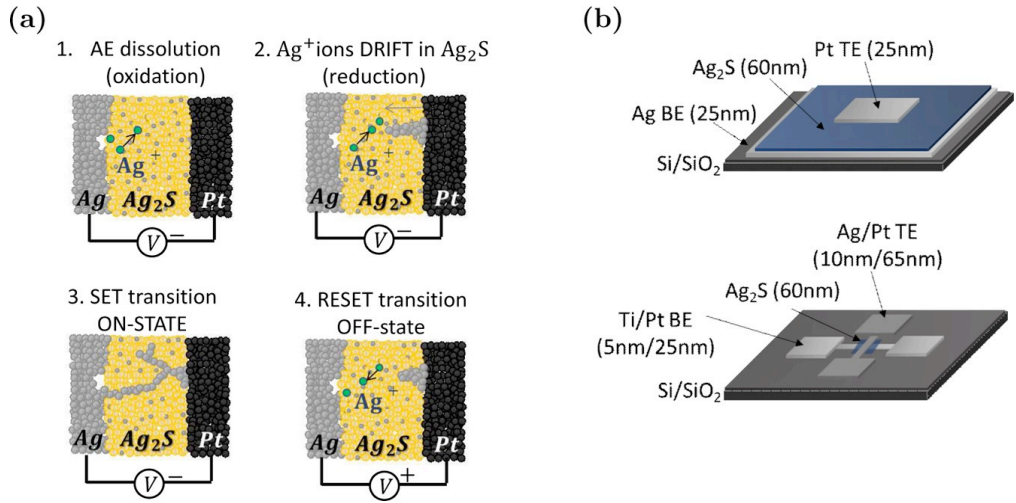

***Figure 1.*** *Filamentary switching. (a) Basic switching mechanism of ECM cells. (b) Device configuration at the millimeter scale (top: 0.1 mm x 0.1 mm active area) and nanometer scale (bottom: 200 nm x 200 nm cross-point active area).*

STP corresponds to a neuronally induced synaptic weight modification that tends to relax toward a resting state, thereby providing activity-dependent signal processing. In LTP, the synaptic weight modification can last for days to months. Thus, LTP provides the information storage capability to the network. Spike timing-dependent plasticity is a variation of Hebb's rule[8,9] that has attracted a lot of attention. Although not involved in all mechanisms of learning, spike timing-dependent plasticity has been demonstrated in various nanoscale memory or memristive devices.[10-17] Other important expressions of plasticity that have been displayed in memristive systems include STP,[18,19] demonstrated based on the volatile memory effect, and the STP to LTP transition,[20-23] displayed in filamentary memory devices in which electrical conductivity is modulated by growth of a conductive filament. Conductive filament growth is induced by the accumulation of electrical stress and leads to an increase in device conductivity. By analogy to long-term memorization processes, which involve the accumulation of short-term effects, and to the idea of reinforcement learning,[24] conductive filament growth has been directly correlated with increased filament stability, corresponding to long-term storage of the conductive state. In these different works, while the strong analogy between biological synapses and nanoscale filamentary memory devices is evidenced, transition between STP and LTP is intrinsic to the material system considered (*i.e.*, ionic species, ionic conductor) and cannot be controlled and tuned during operation.

In this paper, we demonstrate that more complex plastic behaviors can emerge from nanoscale memristive devices, thus allowing a greater number of features to be embedded in a single component and potentially permitting more complex computing systems. By considering more complex filament shapes, such as dendritic metallic paths of different branch densities and widths, we show that the volatile/ nonvolatile regime can be tuned independently, lead- ing to an independent control of STP and LTP. On the basis of the observation of metallic filaments in macroscale electrochemical metallization (ECM) cells, we investigated the growth and stability proper- ties of dendritic filaments. The results were used as a basis for the development of nanoscale solid-state synapses that display independent control of STP and LTP processes *via* spiking excitation and past history modification. When this behavior was interpreted from the framework of the phenomenological modeling developed for synaptic plasticity, the results revealed a strong analogy between our solid-state device and biological synapses. The additional functionality of independent control of STP and LTP could lead to new learning and computing strategies for neuromorphic engineering and artificial neural networks.

## RESULTS AND DISCUSSION

**Ag$_2$S Filamentary Switching.** The basic structure of the synaptic device (Figure 1a) corresponds to a conventional ECM cell, as described by Waser.[25] Inert and reactive Pt and Ag electrodes, respectively, are separated by a Ag$_2$S ionic conductor material (60 nm), which ensures the migration of oxidized Ag$^+$ ions between the electrodes. A positive bias (with a grounded Pt electrode) induces the oxidation of Ag into Ag$^+$ ions at the Ag electrode, the migration of ions from the Ag anode to the Pt cathode, and the reduction of Ag$^+$ ions into Ag filaments across the insulating Ag$_2$S, thereby turning the device from an insulating OFF state to a conductive ON state (SET transition). A negative bias induces the oxidation of Ag from the filament into Ag$^+$ ions and reduction at the Ag electrode, leading to a disruption of the conductive path that turns the device OFF (RESET transition).

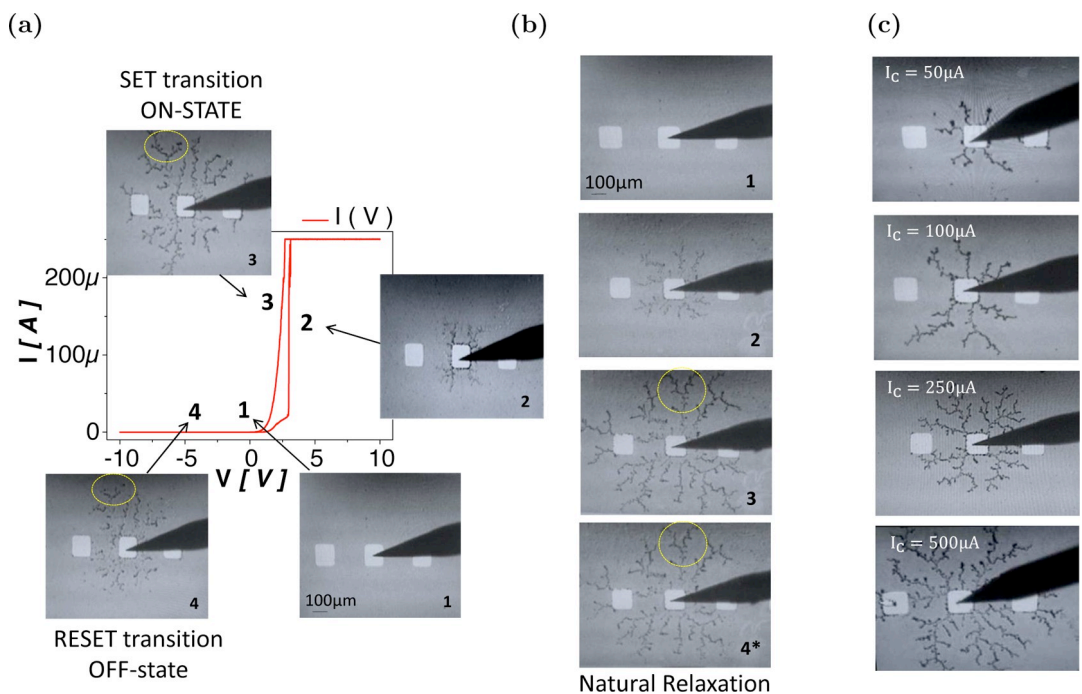

**Figure 2.** Millimeter-scale ECM cell configuration. (a) I-V characteristics and associated optical microscope imaging (0.1mm x 0.1 mm) of filament growth. (b) Natural relaxation of the filament. After a positive SET transition (1-3), the device was kept grounded for 5 min (4*). (c) Relationship between $I_C$ and dendritic expansion/shape.

To gain insight into the filament shape and growth mechanism, we performed optical microscopic imaging during the current-voltage (*I-V*) measurement on millimeter-scale devices with a square-shaped Pt electrode on top of a Ag/$Ag_2S$ substrate (Figure 1b). Consistent with the switching scenario described above, a positive bias induced the formation of Ag dendritic filaments from the Pt cathode toward the Ag anode (SET transition, Figure 2a, snapshot 1 to 3). Application of a negative bias induced a partial destruction of the conducting paths, with remaining filament traces corresponding to preferential paths for subsequent switching (RESET transition, Figure 2a, snapshot 4). After an identical positive SET transition, an intermediate situation was observed, in which the device was kept grounded for 5 min with a slow dissolution of the metallic dendrites (Figure 2a, snap- shot 4*). Such filament relaxation can be attributed to the $Ag^+$ ion diffusion

in the $Ag_2S$ ionic conductor and to the reverse oxidation-reduction process of the Ag filaments.[26]

A second analysis of the filament formation was realized by varying the compliance current ($I_c$) during the SET process. This approach is commonly used in ECM cells to tune the conductance of the ON state and to limit the formation of filaments.[27] If tuning the conductance by limiting the growth of a single filament is considered straightforward (*i.e.,* because the filament diameter corresponds directly to the conductance state), then a more complex picture was obtained for ECM cells that had complex dendritic filament morphologies. Increasing the density or width of the dendritic branch can correspond to an increase of conductance. Because of the resolution of the optical microscope, it was not possible to obtain an accurate assessment of filament diameter. However, we effectively measured a larger filament expansion and dendritic tree density with a larger $I_c$ (Figure 2c). This observation indicates a direct correlation between $I_c$ and the fractal geometry of the dendritic filaments (see Supporting Information, Figure S1). Again, after RESET, the remaining filament traces corresponded to preferential paths for subsequent switching.

Using the previous analysis as a guideline for describing nanoscale filament stability, we implemented the same structure in nanoscale devices consisting of Ag/Pt cross-points with a 200 nm x 200 nm active area separated by $Ag_2S$ (Figure 1b). This device configuration offers the potential for cross-bar integration (cross-point of metallic wires) and for the realization of dense synaptic arrays. Because of the high mobility of the $Ag^+$ ions in the $Ag_2S$ ionic conductor, the device was operated at low voltages, close to the biological electrical potential recorded in neuronal cells during spiking (200 *vs* 80 mV).

As expected, controlling the $I_c$ value during SET transition limited the filament growth and tuned the ON conductance state. ON states at $I_c$ values of 100 nA to 50 μA were strongly volatile, whereas ON states at $I_c$ values above 50 μA were stable, with RESET transition observed at a negative bias (Figure 3a). A linear *I-V* relationship, defining the ON conductance state $G_{ON}$, was obtained in all ON states, indicating that the filaments bridged the gap between the electrodes.

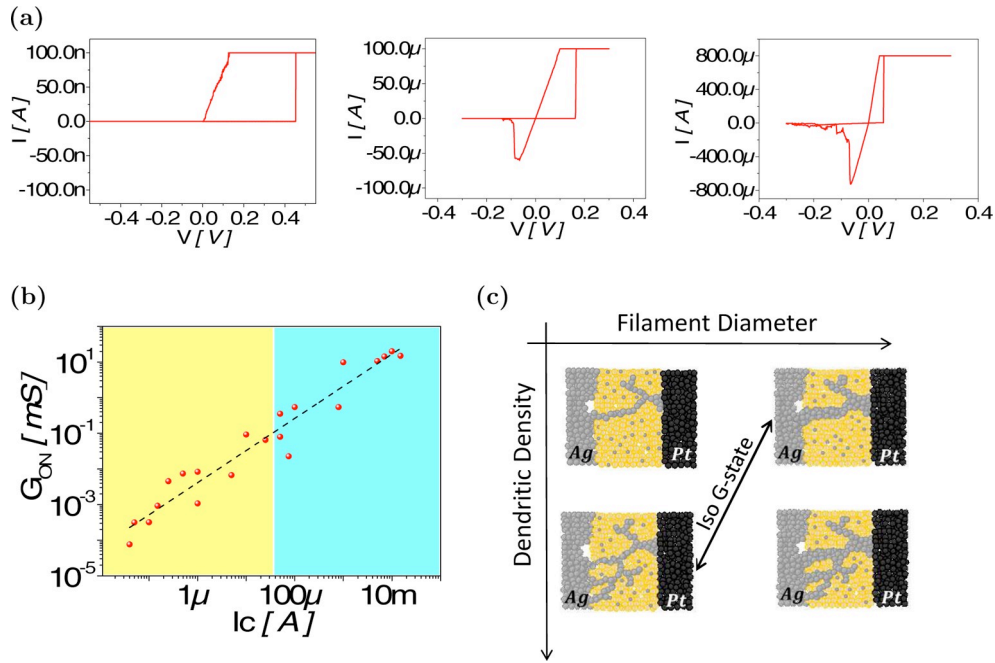

***Figure 3.*** *Nanoscale ECM cell configuration. (a) I-V switching characteristics for different values of the compliance current, $I_C$. When $I_C$ = 100 nA, the ON state is unstable and tends to relax very quickly (OFF transition is not measurable). When $I_C$ = 100 or 800 µA, conventional bipolar switching hysteresis loops are obtained, corresponding to the stable ON state. (b) ON state conductance as a function of $I_C$. Limiting the current during SET limits filament formation. When $I_C$ = 100 nA to 50 µA (region I), the bridging filaments show a high volatility; when $I_C$ > 50 µA (region II), the ON states are stable. (c) Schematic of the proposed scenario describing switching in ECM cells. Both the density and diameter of the dendritic branches can induce an increase in the ON state. The iso-conductance state can be obtained with two different filament configurations.*

Consequently, the large dynamic range of ON states presented in Figure 3b, namely, from high resistance at low $I_c$ (*i.e.*, 1 MΩ at 100 nA, corresponding to a switch- ing power <100 nW), to low resistance at high $I_c$ (*i.e.*, 1$k$Ω at 1 mA, corresponding to a switching power of 300 µW) can be attributed to a modification of the bridging filament morphology, rather than to a modulation of the tunnel barrier length (which is a plausible mechanism in the case of a non-bridging filament).

As a first level of interpretation, the low $I_c$ region can be reasonably described by weak filaments that tend to dissolve very quickly once the voltage is removed. The high $I_c$ region can be considered to correspond to strong bridging filaments with slower relaxation. This effect has been described thermodynamically in Ag filaments[28] as a competition between the surface and volume energies: thin filaments tend to be disrupted because the surface energy is higher than the volume energy, whereas thick filaments tend to stabilize because the volume energy is higher than the surface energy. Such relaxation of the conductive paths has been reported in nanoscale devices[22,23] and was the basis for the implementation of STP and the STP to LTP transition. After the conductive filament forms *via* a strong stimulation, the filaments tend to dissolve and the device relaxes toward its insulating state, leading to STP behavior. Stronger stimulation of the device during the SET transition leads to stronger filaments and higher conductance states with more stable characteristics, resulting in LTP. In this case, the conductance state is correlated directly with the volatility.

Assuming that similar dendritic processes occur at the nanometer and millimeter scales (Figure 2a), we can draw a more complex picture for the interpretation of filament stability. Specifically, the different ON states can be described by dendritic trees, in which the resistance is modulated equally by the density and diameter of the branches. At the nanoscale, the same ON state can be obtained by filaments with dense and thin branches as can be obtained by filaments with less dense and thick branches (Figure 3c). Both configurations should lead to different volatilities, emulating different plasticity properties.

**Synaptic Plasticity Implementation.** To evaluate the plasticity properties of our electronic synapses, we performed pulsed measurements with simplified pulses equivalent to the spike rate-coding scheme observed in biological networks. First, a full SET and RESET cycle was realized by voltage sweeping and limiting the current in the SET transition, with the conditioning loop resulting in an initial OFF state equivalent to Figure 3a. Then, the device was exposed to a train of pulses (5 kHz) with fixed amplitude (0.42 V) and width (100 μs), resulting in potentiation of the device (*i.e.*, conductance increase). Relaxation of the synaptic efficiency was sampled over six decades of time by short read pulses with lower voltage (0.1 V) and short duration (100 μs), to minimize the effect on the relaxation mechanism (Figure 4a). Different excitatory bursts, obtained by varying the number of pulses, were used to modulate the potentiation obtained at the end of the pulse sequence, corresponding to the conductance at the end of a burst of pulses, $G_{max}$. These bursts were fitted by a simple exponential function, ($y = Ae^{-x/t}$, Figure 4b).

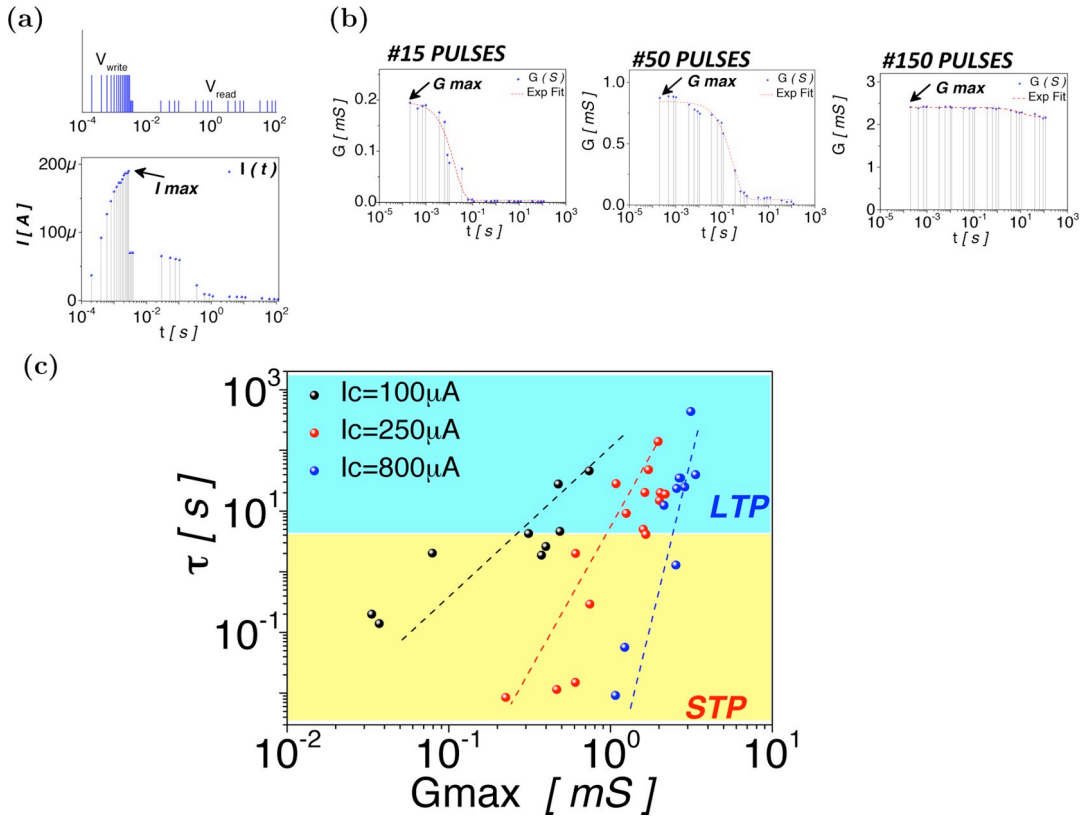

*Figure 4. Implementation of the synaptic plasticity. (a) Protocol for the measurement of pulse relaxation. A burst of pulses at 5 kHz ($V_{write}$ = 0.42 V) induced potentiation. Current relaxation was measured at a lower voltage ($V_{read}$ = 0.1 V) over six decades of time. (b) Measurements of conductance relaxation (blue points) and fitting (red line) on six time decades for different potentiation ($G_{max}$) values, obtained by varying the number of pulses (15, 50, and 150 pulses). Low and high $G_{max}$ values led to STP (complete relaxation over time) and LTP (no relaxation over time), respectively. (c) Relaxation time constant as a function of $I_C$ and conductance state at the end of the burst of pulses, $G_{max}$.*

Consistent with our previous observation that low stability is obtained at a low ON state due to the thinner filaments, we obtained a short relaxation time constant for the lowest ON state. Increasing $G_{max}$ led to a higher time constant and more stable filaments. When we analyzed the evolution of the relaxation time as a function of $G_{max}$ for different $I_c$ values during the conditioning loop (Figure 4c), a second parameter for volatility control emerged. At high $I_c$ values, there was a sharp transition between the low and high time constants. A smoother transition was obtained as $G_{max}$ increased when lower $I_c$ values were used. Another formulation of this result is presented in Figure 5a. If we consider the conductance state 100 s after the end of the excitatory burst, then different transitions from STP (relaxation of the conductance state after 100 s; $G_{max} > G_{100s}$) to LTP (no relaxation of the conductance state after 100 s; $G_{max} = G_{100s}$, blue area in Figure 5a) can be identified as a function of $I_c$. This behavior can be attributed to the combination of two effects. Namely, both $I_c$ and the strength of the excitatory burst (*i.e.*, number of pulses) contribute to the definition of the conductive paths. After the conditioning loop, the device is in its OFF state. Traces for the remaining dendritic branches (defined by $I_c$) correspond to preferential paths for filament formation during the excitatory burst. By analogy with filament formation obtained on millimeter-scale devices, higher $I_c$ should lead to denser dendritic trees. Thus, the first parameter for plasticity tuning is the $I_c$ value used during conditioning. This value controls the average conductance of the filament during switching in pulse mode, by defining the switching path (*i.e.*, dendrite density). The second parameter that controls the STP to LTP transition is the excitation strength (*i.e.*, number of pulses, which controls $G_{max}$). This parameter can be associated with an increase of the branch diameter. These two parameters, the past history of the device through the conditioning loops, and the spiking activity during potentiation can be changed independently of each other to modify the device conductance and the filament volatility. Such mechanism is consistent with simulation and experimental results obtained in.[29] In this work, large filaments are obtained at low surface overpotentials (voltage applied at the electrode/ionic conductor interface) and long switching time while thin filaments results from large surface overpotentials and short switching time. In our case, as the applied voltage is constant, we should consider the voltage redistribution across the full device (*i.e.*, top and bottom interfaces and the ionic conductor's bulk): for low $I_c$ value used during conditioning, the remaining paths correspond to higher bulk's resistivity (low dendritic density) in comparison to high $I_c$ value that leads to denser dendritic paths with lower bulk's resistivity. Consequently, at fixed pulse amplitude, the surface overpotential can be significantly larger in the case of high $I_c$ conditioning loops than in the low $I_c$ regime. In order to reach an equivalent ON state, low $I_c$ conditioning requires a larger switching time (*i.e.*, larger pulse number) leading to large filaments while the high $I_c$ conditioning leads to shorter switching time (*i.e.*, lower pulse number) and thin filaments. To illustrate the improved functionality obtained with our approach, we used the biological model of synaptic plasticity developed by Markram[30] to fit our different synaptic potentiation experiments (Figure 5b). This model describes the excitatory post-synaptic potentiation response produced by a train of presynaptic action potentials (APs). After a number of APs($n$), the postsynaptic current response to the $n+1^{th}$ AP is given by

$$I_{n+1} = A_{SE} R_{n+1} U_{n+1} \tag{1}$$

where the absolute synaptic efficiency, $A_{SE}$, corresponds to the maximum possible synaptic efficiency;

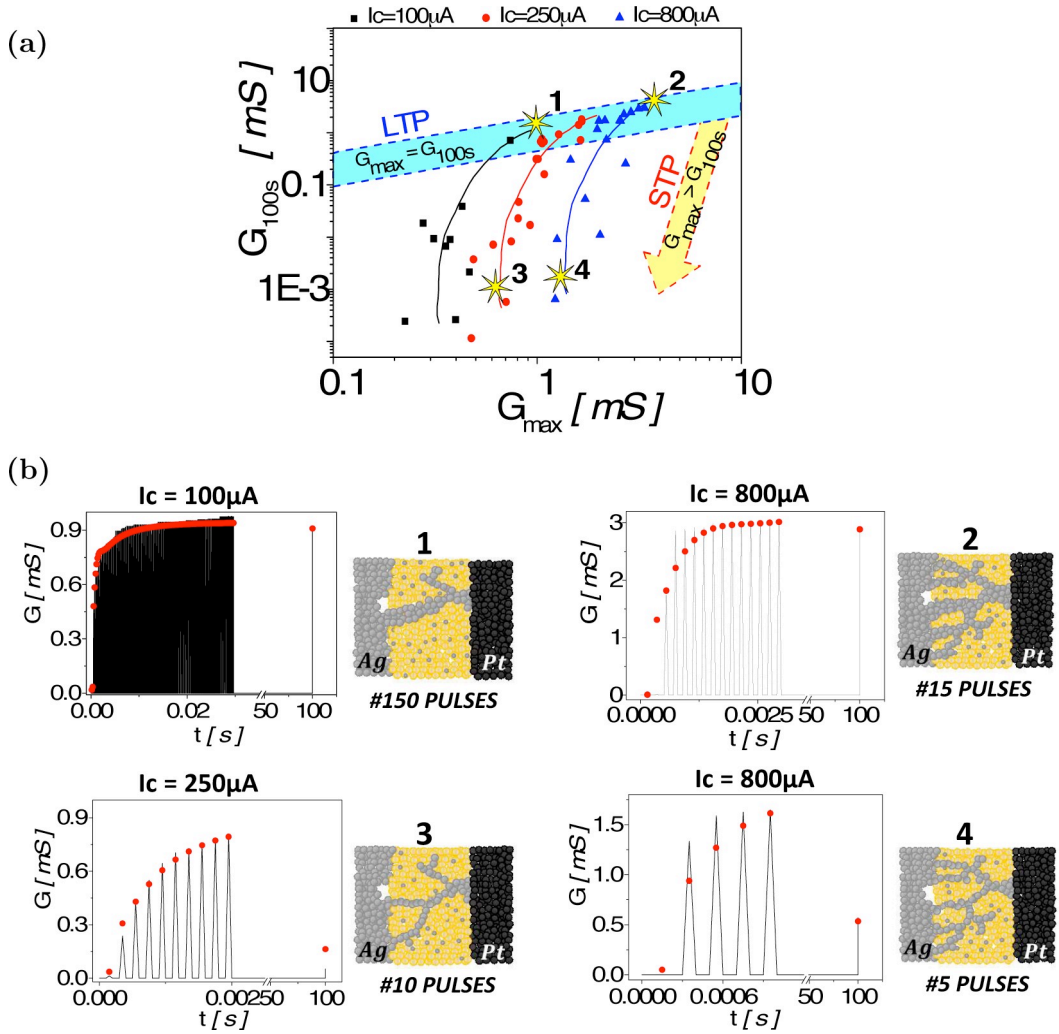

*Figure 5.* Implementation of the synaptic plasticity. (a) After a conditioning loop (full SET and RESET cycle with current compliance, $I_C$), the device is stressed with a burst of spikes, which induce a potentiation from the OFF state to a final conductive ON state, $G_{max}$. Device conductance is measured 100 s after the end of the burst to evaluate the relaxation. Different transitions from STP to LTP are obtained with different conditioning $I_C$ values ($I_C$ = 100, 250, 800 µA). (b) Two examples of LTP (cases 1 and 2) and STP (cases 3 and 4), for the case in which the number of pulses is set as the key plasticity factor and the $I_C$ value is set as the dendritic path definition. The density (through $I_C$) and diameter (through burst excitation) of the dendritic branches can be tuned independently to reproduce various STP/LTP combinations.

*TABLE 1. Fitting Parameters Used for Synaptic Plasticity Modeling*

|  | LTP | | STP | |
| --- | --- | --- | --- | --- |
|  | case 1 | case 2 | case 3 | case 4 |
|  | $I_c = 100\ \mu A$ | $I_c = 800\ \mu A$ | $I_c = 250\ \mu A$ | $I_c = 800\ \mu A$ |
|  | 150 pulses | 15 pulses | 10 pulses | 5 pulses |
|  | $U_{SE} = 0.0279$ | $U_{SE} = 0.0279$ | $U_{SE} = 0.0251$ | $U_{SE} = 0.0279$ |
|  | $A_{SE} = 6$ mS | $A_{SE} = 25$ mS | $A_{SE} = 6.5$ mS | $A_{SE} = 16$ mS |
|  | $\tau_{rec} = 0.0013$ s | $\tau_{rec} = 0.0013$ s | $\tau_{rec} = 0.0010$ s | $\tau_{rec} = 0.0012$ s |
|  | $\tau_{fac} = 11.5500$ s | $\tau_{fac} = 18.5500$ s | $\tau_{fac} = 0.0150$ s | $\tau_{fac} = 1.5500$ s |

the fraction of available synaptic resources, *R*, corresponds to the neurotransmitter resources that are available in the presynaptic connection ($0 < R < 1$); and the utilization of the synaptic efficacy, *U*, corresponds to the amount of neurotransmitter that is released from the pre- to the post-synaptic connection ($0 < U < 1$). Thus, $R_{n+1}$ and $U_{n+1}$ are given by

$$\begin{cases} R_{n+1} = R_n(1-U_{n+1})e^{-\Delta t/\tau_{rec}} + (1-e^{-\Delta t/\tau_{rec}}) \\ U_{n+1} = U_n e^{-\Delta t/\tau_{fac}} + U_{SE}(1-U_n)e^{-\Delta t/\tau_{fac}} \end{cases} \quad (2)$$

The facilitating behavior observed during a burst of spikes is associated with the parameter $U_{SE}$, which is modified with the characteristic time $\tau_{fac}$ and applied to the first AP in a train (i.e., $R_1 = 1 - U_{SE}$). Recovery of the synaptic efficiency (or available neurotransmitters) is associated with the characteristic time τrec. This biological model allows us to reproduce different kinds of plasticity observed in synapses relative to different mechanisms. Plasticity can be controlled through the neurotransmitter dynamics in the presynaptic connection (i.e., recovery of the available neurotransmitters or increase in the neurotransmitter release probability), by the improvement of neurotransmitter detection in the postsynaptic connection or even by a structural modification of the synaptic connection (i.e., increase in the size of a given synapse or the overall number of synapses connecting two neurons). For a detailed review of synaptic plasticity, see refs 31 and 32. Consequently, the synaptic efficiency of a given spike is determined by a combination of parameters that lead to different synaptic responses and expressions of synaptic plasticity.

By accounting for the parameters of the bio-model (eq 2), four different cases may be analyzed as a function of the number of pulses and $I_c$ (Table 1). If we consider experiments 1 and 3 in Figure 5b, the same potentiation (i.e., $G_{max}$ = 0.9 mS) can lead to LTP (case 1 with 150 pulses and $I_c$ = 100 µA) or STP (case 3 with 10 pulses and $I_c$ = 250 µA). The STP to LTP transition is mainly associated with an increase of the facilitating time constant, $\tau_{fac}$. This increase is obtained by increasing the number of pulses during the excitatory burst.

Slightly increasing $I_c$ is mostly represented by an increase in $A_{SE}$. This observation is also evident by comparing case 2 with case 4. The difference in conductance level between cases 1 and 2, which showed qualitatively equivalent LTP responses, is mainly attributed to an increase of $A_{SE}$, from 6 mS (case 1) to 25 mS (case 2). We cannot establish a one-to-one correspondence between biological processes (*e.g.*, neurotransmitter dynamics, structural modifications, *etc.*) and filament growth or relaxation in our experiments be- cause most of the parameters are coupled in both cases. Additional experiments, such as the *in situ* observation of filament shape, would provide more insights in order to formulate of more refined equivalence.

## DISCUSSION

Obtaining the synaptic density has been a major challenge in neuromorphic engineering. From a practical perspective, we believe that developing devices that are more functional (*i.e.*, have properties closer to biological synapses) will allow the construction of more complex systems. In a previous report describing the STP to LTP transition,[14,22] the transition was controlled by a single parameter (*i.e.*, device conductance). Such behavior was proposed as a direct solution for the implementation of the multistore memory model[33] which considers that learning events contribute to the formation of short-term memory (where memory is used in the sense of psychology) before being transferred into long-term memory (STM/LTM transition). If a direct equivalence between STP/LTP and STM/LTM is not straightforward, it seems realistic to consider synaptic plasticity as a key element in the formation of memory. The device presented in this paper features a tunable STP/LTP transition that could be a key parameter for defining the appropriate activity threshold that determines when information storage needs to be moved from a short-term to a long-term regime, or, in other words, how long an information needs to be sustained (*i.e.*, how long the device will remain in its ON state). Additionally, if STP/LTP transition is only controlled by the device's conductance, synaptic weight modifi- cation and STP/LTP transition cannot be uncorrelated. We argue that the rate-coding property obtained in the STP regime, as observed in the facilitation of synaptic signal transmission during a high frequency burst of spikes and the subsequent relaxation at lower frequencies, disappears once the device enters into its LTP regime and, thus, becomes a linear resistor. From a circuit perspective, if we consider a simple integrate- and-fire neuron associated with linear synapses, the node (neuron and synapses) is equivalent to a simple linear filter (if the variable is the average spiking rate). The node is a nonlinear filter in the STP regime with frequency-dependent synaptic conductance. The overall network functionality is reduced when learning moves synapses from their STP to their LTP domain. An interesting property offered by the presented devices in order to preserved such rate coding functionality is to allow for weight modification through the control of the $A_{SE}$ parameter while maintaining the frequency dependent response by keeping the device into its short-term regime (see case 3 and 4, Figure 5). For the device presented in this paper, learning can be realized by modifying the dendritic filament density and increasing the $A_{SE}$ during the conditioning procedure. The frequency coding property can be ensured by controlling the filament diameter and relaxation.

Finally, the activity dependent STP/LTP transition and synaptic weight modification in this work is only obtained as a function of the input frequency, thus corresponding to the pre-neuron activity. Such mechanism is defined in biology as a facilitating synapse. A complementary mechanism, that cannot be reproduce with our system, is the depressing synapse (*i.e.*, decrease of the synaptic weight when pre-neuron activity increase). In order to implement practical learning systems, these results will have to be extended to hebbian learning strategies in which weight modification is dependent on both pre- and post-neuron activity. Among the different hebbian learning strategies considered to date, STDP has attracted a large attention. One implementation of such learning protocol is based on overlapping pulses (spike timing difference between pre- and post-neuron is then encoded as a voltage drop across the device). Figure S3 presents similar results to Figure 5 when voltage is used as a key plasticity factor instead of spiking frequency that should allow for STDP realization. While not measured in this paper, one interesting future direction would be to add to previously reported STDP results obtained on nonvolatile systems[11] the STP/LTP capacity in order to demonstrate neuromorphic circuits with richer dynamical behaviors.

## CONCLUSIONS

We report a single synaptic device that highly resembles its biological counterpart, opening the field to more complex neuromorphic systems. Biological synaptic plasticity has been successfully implemented in our nanoscale memristive device by considering the filament stability of ECM cells, in terms of competition between the density and diameter of the dendritic branches. STP and LTP regimes can be controlled by tuning the device volatility. The first parameter for plasticity tuning, $I_c$, is used during conditioning and controls the average conductance of the filament during switching in pulse mode. The second parameter handles the STP to LTP transition and corresponds to the excitation strength (number of pulses), which controls $G_{max}$. The second parameter can be associated with an increase of the branch diameter. These two parameters can be tuned independently of each other to modify the device conductance and filament volatility.

Future work should investigate how such synaptic properties can be advantageous for large-scale neuromorphic circuits. To improve the efficiency of future bio-inspired computing systems, interdisciplinary re- search is needed to obtain a better understanding of the contributions of STP and LTP mechanisms to memory construction and spike-coding information processing.

## METHODS

**Device Structure and Fabrication.** To fabricate the millimeter-scale Ag/Ag$_2$S/Pt device (Figure 1b), a 25 nm Ag bottom electrode was deposited by direct current magnetron sputtering onto the cleaned surface of *p-type* silicon, which was covered with thermally grown 200 nm thick SiO$_2$ at room temperature. A thin film of Ag$_2$S (60 nm) was deposited by thermal evaporation onto the full substrate. Finally, a Pt top electrode, with a thickness of 25 nm and electrode size of 1 mm, was deposited on the Ag$_2$S layer by using a shadow mask and direct current magnetron sputtering. To fabricate the nanoscale device, a 5 nm/25 nm Ti/Pt bottom electrode was deposited at room temperature on thermally grown $SiO_2$ (200 nm) and patterned *via* lift-off and electron beam lithography (EBL). A thin film of $Ag_2S$ (60 nm) was deposited by thermal evaporation and patterned *via* lift-off and EBL. Finally, a 10 nm/70 nm Ag/Pt top electrode was deposited on the $Ag_2S$ by direct current magnetron sputtering and patterned *via* lift-off and EBL.

**Characterization of the Switching Dynamics.** The electrical characteristics of the device, the waveform design of the pulses at fixed amplitude ($V_{write}$ = 0.4 V and $V_{read}$ = 0.1 V) and width (w = 100 μs), and the pulse measurements were obtained by using a semiconductor device analyzer (B1500A, Agilent) and a waveform function generator (WGFMU B1530A, Agilent), which were piloted in remote mode by VISUAL STUDIO. For all measurements, the device electrodes were contacted with a micro-manipulator probe-station (Suss Microtec PM-5), and the Pt electrode was grounded. Conditioning loops, characterized by a full SET and RESET cycle, were realized by sweeping voltage and limiting current in the SET transition.

*Acknowledgment.* The authors thank D. Guerin, F. Vaurette, S. Lenfant for technical expertises and D. Querlioz for careful reading of the manuscript. This work was supported by the ANR DINAMO project (no. ANR-12-PDOC-0027-01).

*Supporting Information Available:* Fractal analysis of the dendritic filaments and implementation of the synaptic plasticity with the pulse amplitude as the key plasticity factor. This material is available free of charge *via* the Internet at http://pubs.acs.org.

# Filamentary Switching: synaptic plasticity through device volatility

S. La Barbera, D. Vuillaume, and F. Alibart

*Institut d'Electronique, Microelectronique et Nanotechnologies, UMR-CNRS 8520*

E-mail: fabien.alibart@iemn.univ-lille1.fr

## Supporting Information

An useful analysis to investigate the filamentary shape can be done by exploiting the fractal geometry. Indeed, fractal analysis provides a useful framework for complex pattern description that is not well described by common euclidean measures such as diameter or length. By properly choosing a region of an optical image (*60px × 110px*) and by converting it in a binary image (Figure S1) it is possible to estimate the fractal dimension (*D*) and its lacunarity (*λ*). These calculations are made through *ImageJ*, a software that allows to count the number of boxes of an increasing size needed to cover a one pixel binary object boundary and implements the fractal method as described in ref. [1]. A plot is generated with the log of size on the X-axis and the log of count on the Y-axis and the data is fitted with a straight line.

The slope (*S*) of the line is the negative of the fractal dimension, i.e., $D = -slope$. The lacunarity (*λ*) can be defined as the measure of the fractal structural variation or fractal texture. It is calcuated from the standard deviation ($\sigma$), and mean ($\mu$), for pixels per box, i.e. $\lambda = (\sigma/\mu)^2$. Thus, *D* and *λ* work together to characterize complex patterns extracted from digital images.

Figure S2 presents the evolution of fractal parameters as a function of $I_c$ during SET. A clear correlation and anti-correlation with $I_c$ was obtained for $\lambda$ and $D$, respectively. These latter parameters do not provide a direct description of dendritic branches density and width, but such evolution is in agreement with the proposed scenario described in the paper. Further investigation will be carried to exploit fractal geometry description of filamentary switching.

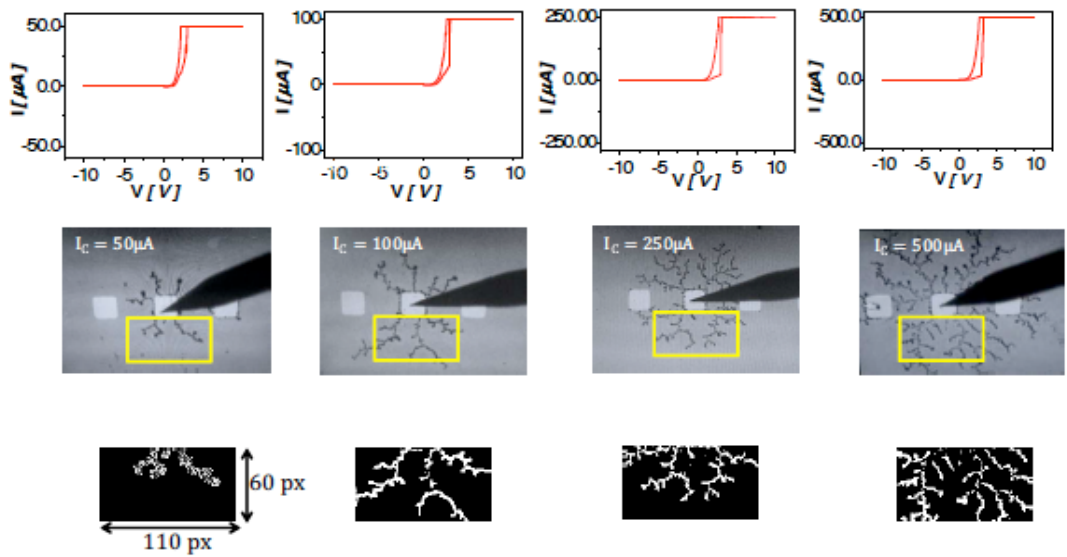

*Figure S1:* Filamentary switching analysis through fractal geometry. Conditioning loops for $I_c$ = 50, 100, 250 and 500µA, correspective optical microscope imaging (1mm x 1mm) of the filament growth and binary images of the selected yellow region (60px × 110px).

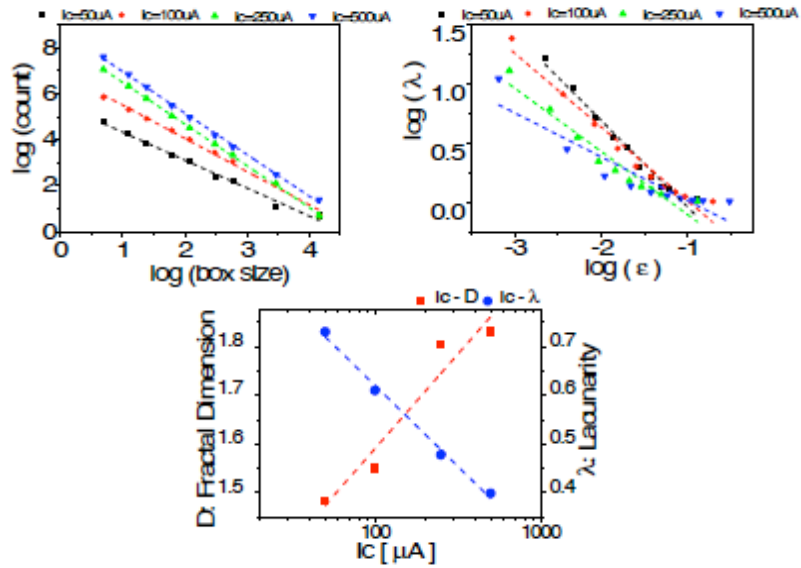

*Figure S2:* *Filamentary switching analysis through fractal geometry. Fractal dimension D and lacunarity λ parameter calculation and relation with $I_c$.*

Synaptic plasticity can be implemented by different burst configurations that module the potentiation obtained at the end of the pulse sequence (corresponding to the conductance at the end of a burst of pulses, $G_{max}$): stronger potentiation can be obtained by increasing the number of pulses in the burst (as shown in the main text) or the pulse amplitude (Figure S3), parameter that has been proposed for Spike Timing Dependent Plasticity (STDP) based on overlapping pulses.

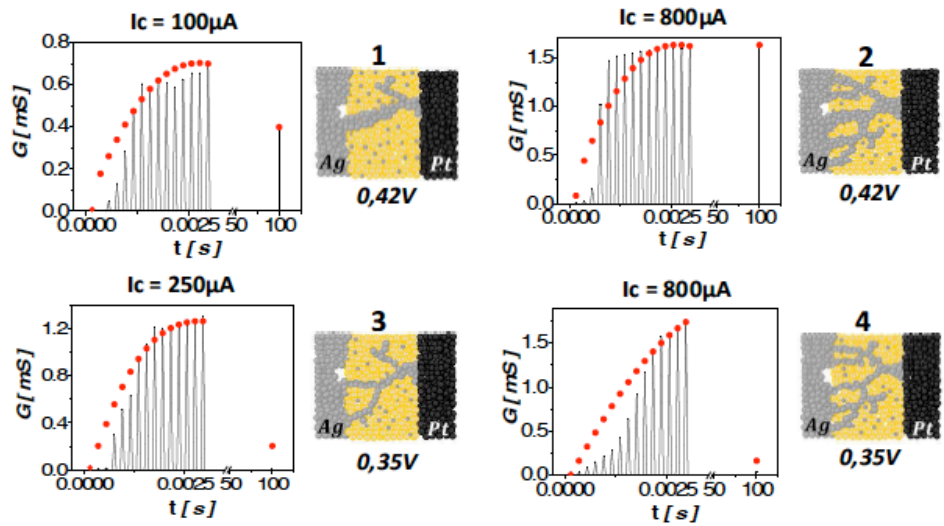

*Figure S3*: Implementation of the synaptic plasticity. By using the the pulse amplitude as plasticity key factor. Two examples of LTP (case 1 and 2) and STP (case 3 and 4) are obtained. Both dendritic branches density and dendrites diameter can be tune independently to reproduce various STP/LTP combinations